# Evidence for a universal Fermi-liquid scattering rate throughout the phase diagram of the copper-oxide superconductors

N. Barišić[1,2,3], M. K. Chan[2,a], M. J. Veit[2,b], C. J. Dorow[2,c], Y. Ge[2,d], Y. Li[2], W. Tabis[2,e], Y. Tang[2], G. Yu[2], X. Zhao[2,4], and M. Greven[2]

[1] Institute of Solid State Physics, TU Wien, 1040 Vienna, Austria

[2] School of Physics and Astronomy, University of Minnesota, Minneapolis, MN 55455, USA

[3] Department of Physics, Faculty of Science, University of Zagreb, HR–10000 Zagreb, Croatia

[4] State Key Lab of Inorganic Synthesis and Preparative Chemistry, College of Chemistry, Jilin University, Changchun 130012, China

Present addresses:

[a] National High Magnetic Field Laboratory, Los Alamos National Laboratory, Los Alamos, NM 87545, USA

[b] Department of Applied Physics, Stanford University, Stanford, CA 94305, USA

[c] Department of Physics, University of California, San Diego, CA 92093, USA

[d] Department of Physics, Penn State University, University Park, PA 16802, USA

[e] Laboratoire National des Champs Magnétiques Intenses, UPR 3228, (CNRS-INSA-UJF-UPS), Toulouse 31400, France

# ABSTRACT

The phase diagram of the cuprate superconductors continues to pose formidable scientific challenges. While these materials are typically viewed as doped Mott insulators, it is well known that they are Fermi liquids at high hole-dopant concentrations. It was recently demonstrated that at moderate doping, in the pseudogap region of the phase diagram, the charge carriers are also best described as Fermi liquid. Nevertheless, the relationship between the two Fermi-liquid regions and the nature of the strange-metal state at intermediate doping have remained unsolved. Here we show in the case of the model cuprate superconductor $HgBa_2CuO_{4+\delta}$ that the scattering rate measured by the cotangent of the Hall angle remains quadratic in temperature across the pseudogap temperature, upon entering the strange-metal state, and that it is doping-independent below optimal doping. Analysis of the published results for other cuprates reveals that this behavior is universal throughout the entire phase diagram and points to a pervasive Fermi-liquid transport scattering rate. We argue that these observations can be reconciled with other data upon considering the possibility that the pseudogap phenomenon signifies the completion of the gradual, non-uniform localization of one hole per planar $CuO_2$ unit upon cooling.

# I. INTRODUCTION

The cuprates exhibit a complex phenomenology that is often argued to be incompatible with Landau's highly successful Fermi-liquid (FL) description of conventional metals [1,2], except at very high hole-dopant concentrations $p$ [3,4]. In the undoped state, these materials are Mott insulators with strong antiferromagnetic correlations. The approximately linear temperature dependence of the planar dc-resistivity ($\rho \propto T$) in the so-called strange-metal (SM) region of the phase diagram (Fig. 1a) is observed from about 10 K to 1000 K [5] and can reach rather large values in excess of 1 mΩcm. This is a seemingly unphysical situation wherein the electronic mean-free path appears to be shorter than the lattice constant, and hence the Ioffe-Regel limit is crossed. Consequently, the conventional theoretical description of metals based on quasiparticles with well-defined crystal momenta that experience occasional scattering events is often argued to be inadequate [1]. At the hole concentration $p^* \approx 0.19$, just above the optimal doping level at which the superconducting (SC) transition temperature ($T_c$) is maximized, the SC state directly evolves from the peculiar SM state. The intervening region of the phase diagram at lower doping is marked by the pseudogap (PG) phenomenon below $T^*(p)$ and by evidence for a phase transition [6,7,8,9]. These and other experimental observations have been argued to be the result of an underlying quantum critical point at $p^*$ with associated fluctuations that give rise to the extended $T$-linear resistive behavior and to the superconductivity [1,2]. Nevertheless, despite these peculiarities, there also have been suggestions that, fundamentally, FL theory should prevail [10-14], and that the Ioffe-Regel condition is not violated [14,15]. In fact, there has been a recent revival of Fermi-liquid concepts in correlated systems in general, including considerations of a FL scattering rate that extends to high temperatures (Appendix A).

HgBa$_2$CuO$_{4+\delta}$ (Hg1201) may be viewed as a model compound due to its relative structural simplicity (tetragonal symmetry, one CuO$_2$ layer per formula unit, no CuO chains), negligible residual resistivity and large optimal $T_c$ of nearly 100 K [16]. Below $T^{**}$ ($T^{**} < T^*$; Fig. 1a), the planar resistivity of Hg1201 exhibits quadratic temperature dependence, $\rho \propto T^2$, the behavior characteristic of a FL [14]. This finding motivated optical conductivity measurements that yielded the quadratic frequency dependence and the temperature-frequency scaling of the optical scattering rate close to that predicted theoretically for the umklapp Fermi liquid [17]. Furthermore, at temperatures below $T^{**}$, the magnetoresistance of Hg1201 was found to obey Kohler's rule, a property of conventional metals [18].

In contrast to previous approaches that extended ideas developed for the SM phase to the PG phase, we start here from the now well-documented PG/FL state of Hg1201 [14,17,18] and find that the cotangent of the Hall angle, and hence the transport scattering rate remains quadratic in temperature even in the SM phase and that it is independent of doping in the range $0.05 < p < 0.14$. We then demonstrate that this important property is in fact doping and compound independent throughout the entire phase diagram, and hence universal. Although the magnitude of the quadratic temperature dependence of the cotangent of the Hall angle may exhibit compound-specific deviations from the underlying universal value, we demonstrate for the cuprate La$_{2-x}$Sr$_x$CuO$_{4+\delta}$ (where this effect is the largest due to the presence of a Lifshitz transition) that these deviations can be understood as a breakdown of the effective-mass approximation rather than of the Fermi-liquid concept. Based on these new insights, we propose a fundamentally new understanding of the cuprates: the seemingly disconnected PG/FL [14,17,18] and overdoped FL [3,4] regions are in some sense connected, and the notion that the carrier density in the SM region is independent of temperature has to be abandoned.

## II. TRANSPORT MEASUREMENTS

As noted, these conclusions follow from the behavior of the cotangent of the Hall angle, $\cot(\Theta_H)$, which is the ratio of planar resistivity, $\rho$, to Hall resistivity, $\rho_H = HR_H$, where $H$ is the magnetic field and $R_H$ is the Hall coefficient. Within the simple effective-mass approximation that assumes a parabolic band, $\rho = m^*/(ne^2\tau)$ and $R_H = 1/(ne)$, where $m^*$ is the effective mass and $n$ is the carrier density, so that $\cot(\Theta_H) = \rho/\rho_H \propto m^*/\tau$ is independent of the carrier density and essentially a direct measure of the scattering rate $1/\tau$.

Figure 2 shows the temperature dependences of the resistivity and Hall coefficient for Hg1201 up to 400 K (see Appendix B for experimental details). The resistivity exhibits the well-known (approximate) linear behavior at temperatures above $T^*$ and a quadratic dependence in the PG/FL state below $T^{**}$. One of our key findings for this model cuprate is that $\cot(\Theta_H) \propto T^2$ not only in the PG/FL phase, where $1/\tau \propto T^2$ has been established [14,18], but that this quadratic temperature dependence remains unaffected upon crossing $T^{**}$ and $T^*$, and moreover is independent of doping. This is demonstrated in Fig. 2d for three doping levels and summarized in Fig. 3 along with matching results for five additional doping levels (see also Appendix C). Since there exist no noticeable changes in behavior at $T^{**}$ and $T^*$, the simplest and most natural interpretation is that $\cot(\Theta_H)$ continues to measure the scattering rate at temperatures above $T^{**}$, which would imply that the effective mass is doping and temperature independent, consistent with other experimental observations [14,19-23]. A seemingly less likely possibility is a coincidental compensation effect across distinct parts of the phase diagram involving a scattering rate and effective mass that are both temperature- and doping-dependent in a manner that conspires to give rise to the simple and robust $\cot(\Theta_H) \propto T^2$ behavior.

### III. COMPARISON WITH OTHER SINGLE-LAYER CUPRATES

Motivated by these observations for Hg1201, we compare in Fig. 3 our data for $\cot(\Theta_H)$ to already published results for the thoroughly investigated cuprates $La_{2-x}Sr_xCuO_4$ (LSCO) [24] and $Tl_2Ba_2CuO_{6+\delta}$ (Tl2201) [25]. The three compounds have in common that they all feature a single $CuO_2$ layer per structural unit. The LSCO data correspond to the PG region of the phase diagram ($T < T^*$) and extend to the non-SC state at low doping, whereas the Tl2201 results correspond to high doping levels, where FL behavior has been firmly established [1-4]. The combined data span nearly the entire phase diagram. All three compounds exhibit a quadratic temperature dependence, $\cot(\Theta_H) = C_0 + C_2 T^2$. The compound/sample specific residual term $C_0$ is related to disorder [26] and effectively absent in Hg1201. Remarkably, within the measurement error, all three cuprates (except LSCO for $p > 0.08$; see below) exhibit the same value of $C_2$ (see also Appendix D). Since there is good evidence that the effective mass is essentially independent of compound and doping [14,19-23], this indicates that the transport scattering rate is universal across the cuprate phase diagram (Fig. 1b).

This remarkable result is highlighted in Fig. 4, which shows representative data for the temperature dependences of $\cot(\Theta_H) - C_0$ on a log-log scale for the three single-layer cuprates LSCO, Tl2201 and Hg1201. The respective data correspond to different parts of the phase diagram, and range from highly underdoped (LSCO, $p = 0.07$) to highly overdoped (Tl2201, $p \approx 0.26$). For Hg1201 ($p \approx 0.145$), they span different regions (from $T < T^{**}$ to $T > T^*$), without any noticeable change in behavior. Clearly, the slope ($C_2$) is the same in all cases.

## IV. IMPLICATIONS FOR THE FERMI SURFACE AT LOW TEMPERATURES

The main difficulty in understanding the resistivity, Hall coefficient and other electronic properties of the cuprates stems from the rather unusual Fermi-surface evolution with temperature [27] and doping [28] (see also Appendix D). The simple universal quadratic temperature dependence of the Hall angle established here provides a crucial new constraint. Accordingly, two Fermi-surface changes with temperature/doping should be distinguished and carefully considered.

First, we consider the character of the Fermi surface at temperatures below $T^{**}$. The most prominent feature in the PG phase is the antinodal gap at the Fermi level. This partial gap does not appear to affect the near-nodal parts of the Fermi surface (arcs), but rather determines their length [27]. The carrier density in the PG phase, as deduced from dc-resistivity [14] and optical conductivity [17] measurements is proportional to the hole concentration $p$, and the observation of FL charge transport [14,17,18] suggests that the arcs must be well formed below $T^{**}$ [14]. Indeed, the Hall density $n_H = 1/(eR_H)$ is approximately constant for $T_c < T < T^{**}$. For Hg1201, this is best seen for the $T_c = 71$ K sample (Fig. 2c), whereas for LSCO this observation extends over a wider temperature/doping range [24]. Importantly, $n_H$ yields a carrier density that corresponds quite well to the nominal hole density, $n_H \approx p/V$, where $V$ is the unit cell volume. For LSCO, this is the case up to about $p = x = 0.08$ [24]. For Hg1201, this holds for the entire doping range of the present study, apart from a systematic offset of about 20% that can be ascribed to the uncertainty in determining the sample geometry and/or in the absolute value of $p$ in cuprates in which hole carriers are introduced into the $CuO_2$ planes via excess oxygen atoms in the interstitial layers (Appendix C). Deep in the PG/FL region of Hg1201, where the resistivity exhibits a simple quadratic temperature dependence, there remains a relatively weak temperature dependence of $R_H$. In contrast to the resistivity ($\rho = 1/\sigma_{xx}$), the Hall coefficient ($R_H = \sigma_{xy}/(H\sigma_{xx}\sigma_{yy})$, where σ is the conductivity tensor) is sensitive (through $\sigma_{xy}$) to the curvature and the details of the (exact) termination of the arcs. The weak temperature dependence of $R_H$ between $T_c$ and $T^{**}$ might therefore be related to the subtleties of the arc-tip formation. Alternatively, it might be associated with the appearance of SC traces or with charge-density-wave correlations [29].

Second, with increasing hole concentration, the cuprates undergo a zero-temperature topological change from Fermi arcs to a closed Fermi surface in a non-universal, compound-specific fashion [28]. In the general case of a non-parabolic electronic band, the off-diagonal terms of the conductivity tensor need to be taken into account, as noted in the case of LSCO [30]. For LSCO, the observed deviation from $n_H = p/V$, for $p > 0.08$ and $T < T^{**}$, and the apparent divergence of $n_H$ at $p \approx 0.22$ are related to changes in the curvature of the Fermi surface [28] (Appendix D). The Hall coefficient therefore ceases to be a proper measure of the carrier density in this doping range, and the observed behavior is the result of the failure of the effective-mass approximation rather than of the applicability of the FL concept [30]. Indeed, for LSCO at $p > 0.08$, $\cot(\Theta_H)$ remains quadratic in temperature [24], but $C_2$ starts to increase (Fig. 3b) concomitantly with the change in curvature of the underlying Fermi surface. We therefore expect the Hall effect to systematically overestimate the carrier density in this doping range. We can test this hypothesis by correcting the Hall data based on the assumption that the true doping level for $La_{2-x}Sr_xCuO_4$ remains $p = x$ at temperatures below $T^{**}$ (Appendix D), consistent with the dc resistivity [14] and optical conductivity [17] results. As shown in Fig. 3b, the corrected values of $C_2$ indeed agree with the universal value. Evidently, as also demonstrated in Fig. 3b, no such correction is necessary for Hg1201 and Tl2201 in the studied doping ranges. We note that comparable values of $C_2$ have also been obtained for multi-layer cuprates and that, similar to LSCO, there exist systematic deviations from the underlying universal behavior near optimal doping (Appendix E).

These observations provide new insight into the charge transport throughout the entire phase diagram. Since the universal magnitude and temperature dependence of the scattering rate directly connects the PG/FL, SM and overdoped FL regions without any changes at $T^*$ and $T^{**}$, the archetypal linear temperature dependence of the resistivity may not reflect a linear scattering rate, contrary to common belief [1,2]. Instead, the $\rho \propto T$ behavior appears to be the result of a temperature-dependent carrier density, consistent with the well-known approximate $1/R_H \propto T$ behavior for $T > T^*$ [13,24,31].

## V. FERMI SURFACE AND INTRINSIC INHOMOGENEITY

The canonical Landau Fermi liquid features well-defined low-energy quasiparticles whose density does not depend on temperature and satisfies Luttinger's theorem at zero temperature, namely that the particle density enclosed by the Fermi surface equals the total electron density [4,32]. The situation is clearly more complex in the case of the cuprates, which seemingly lack a continuous Fermi surface in the PG phase. Nevertheless, there exist well-defined "nodal" quasiparticles [33,34], Fermi-liquid behavior is observed [14,17,18] and, as discussed, the Hall coefficient in compounds such as Hg1201 and LSCO is a good measure of the carrier density deep in the PG/FL region of the phase diagram [24]. A simple estimate of the carrier density from the arc length in the PG/FL regime indeed suggests that the density is low and increases with doping ($\approx p$) [14]. More rigorous theoretical modeling yields the same conclusion [13,30]. The Hall data, including the present result, are overall consistent with an evolution from $p$ carriers that reside on arcs at temperatures below $T^{**}$ toward a full Fermi surface of size $1 + p$ not only at large doping levels [4], but also at high temperatures [24,31] (see also Appendixes F and G). Although some photoemission experiments indicate that the full Fermi surface may already be recovered at $T^*$ [35], we note that the density of states at the Fermi level is not a measure of the carrier density.

Our findings also call for a reinterpretation of experimental results for the SM state. In particular, it is of interest to consider optical conductivity and photoemission results along the lines discussed in the Appendix G. The gradual delocalization with increasing doping and/or temperature of one hole per unit cell might be the result of the dual covalent and ionic character of the cuprates [10,12]. In order to better understand how such an unusual behavior might arise, we suggest that it is important to recognize that the cuprates feature considerable inherent structural and hence electronic disorder [36-39]. Bulk nuclear magnetic resonance measurements reveal significant spatial inhomogeneity even for simple tetragonal Hg1201 [40], yet they also indicate the simultaneous presence of local PG and FL electronic components in the $CuO_2$ planes [41]. Upon increasing the temperature above $T^{**}$, the physical properties are affected by the closing of the unconventional PG. In real space, the PG manifests itself as inhomogeneous sub-nanoscale electronic gaps, as detected by scanning tunneling microscopy/spectroscopy [36,42]. Real-space gaps exist even at temperatures well above $T^*$, and they were found to close at temperatures proportional to the local gap magnitude [43]. It has been proposed [44] that ~ 100% and ~ 50% PG coverage of the $CuO_2$ planes corresponds to two characteristic temperatures that match $T^{**}$ and $T^*$, respectively. In this picture, $T^*$ signifies a percolation transition. At a qualitative level, this is consistent with the evidence from polarized neutron diffraction [6,7],

Kerr effect [8], and ultrasound [9] experiments for a phase transition at $T^*$ and with the concomitant absence of a specific heat anomaly [45].

For correlated metals, it was noted that well-defined quasiparticle excitations may survive at temperatures well above the range of validity of FL theory as signified by the breakdown of a quadratic temperature dependence of the resistivity, before the Ioffe-Regel limit is reached [46]. However, the fact that the scattering rate is nearly unchanged across the cuprate phase diagram (at least up to 400 K) suggests that the effective Fermi energy at low doping is essentially the same as in the overdoped region ($E_F \sim 1$ eV) [47]. This is consistent with the observation of a large, doping-independent Fermi velocity in photoemission studies [33] and with the notion that the effective mass is doping independent and (nearly) compound independent [14,19-23]. A large effective Fermi energy ($E_F \gg k_BT$) would also explain why FL-like behavior extends to rather high temperatures. Assuming a simple two-dimensional parabolic band, it would appear that the Ioffe-Regel limit ($k_Fl = hc/(\rho e^2) \sim 1$, where $c$ is the interlayer distance) is approached at 400 K in our most underdoped Hg1201 samples [14]. However, this analysis does not take into account the fact that parts of the Fermi surface are removed by the PG and that the mobile carrier density is overestimated, which gives the false impression that the Ioffe-Regel limit is crossed [14,15]. In fact, the room-temperature mobility, $\mu = (H\cot(\Theta_H))^{-1} \approx 10$ cm$^2$V$^{-1}$s$^{-1}$, is not unusual and comparable to that of ordinary metals such as Aluminum [48]. The situation appears to be more complex in underdoped compounds such as LSCO that exhibit a considerable non-universal resistivity contribution characterized by a logarithmic low-temperature upturn [18,23], although the underlying FL transport behavior persists to very low hole-dopant concentrations, as seen from the universal temperature dependences of the resistivity [14,24] and of cot($\Theta_H$). This additional complexity is similar to that in certain inhomogeneous conventional metals [18,23,49]. We note that it was recently demonstrated that FL transport behavior (with nearly the same value of $C_2$) prevails even deep in the AF phase of the electron-doped cuprates, where the non-universal resistivity upturn is particularly prominent [50].

## VI. CONCLUDING REMARKS

Our findings point to the need for a paradigm shift regarding the cuprate phase diagram. The observation that $\cot(\Theta_H) \propto T^2$ in the SM region was first made in the early days of cuprate research and, in fact, a Fermi-liquid interpretation was given in the early 1990s [10]. However, this interpretation of the charge transport has not been given much attention since crucial experimental facts were not known. In conjunction with the recent demonstration of FL charge transport in the PG region [14,17,18], the results for Hg1201 reported here unambiguously demonstrate a profound connection between the PG/FL and SM states, as $\cot(\Theta_H) = C_2 T^2$ remains unaffected upon crossing $T^{**}$ and $T^*$. Furthermore, we demonstrate in Figs. 3 & 4 that this behavior is, in its essence, doping and (nearly) compound independent, and thus universal, which establishes a direct connection between the FL/PG region and the well-accepted FL state on the overdoped side of the phase diagram.

The universal temperature dependence of $\cot(\Theta_H)$ and, in particular, the de facto universal value of $C_2$ provides an unprecedented constraint on any physical interpretation. To our knowledge, no prior interpretation except the Fermi liquid discussed here is consistent with the above experimental facts. Whereas some theoretical models seem to capture the experimental observations made in the overdoped region, where with decreasing doping the planar resistivity gradually evolves from quadratic to linear temperature dependence, we emphasize that, at a fixed doping level, the resistivity in the model compound Hg1201 switches from purely $T^2$ (PG/FL) to purely $T$ (SM) without any change in $C_2$. This fact alone excludes interpretations of the transport properties based on an anisotropic scattering rate (e.g., [25,26]). Namely, it would seem highly unphysical that in distinct parts of the phase diagram ($T < T^{**}$ versus $T > T^*$) the scattering rates are qualitatively different (or correspond to different parts of the Fermi surface) without any change in $\cot(\Theta_H) = C_2 T^2$. Moreover, the fact that $\cot(\Theta_H)$ is doping and compound independent provides a profound connection among all the regions of the phase diagram, and thereby excludes narrow theoretical interpretations based on the behavior of any particular compound in a specific doping and temperature range.

One pivotal goal of our work is to separate underlying fundamental properties of the curpates from those that are compound specific. Only once the former are firmly established is it feasible to address the latter. In this regard, it is important to note that compounds such as $Bi_2(Sr,La)_2CuO_{6+\delta}$ (Bi2201) and $Bi_2Sr_2CaCu_2O_{8+\delta}$ (Bi2212) do not show the underlying quadratic temperature dependence of the planar resistivity in the PG/FL regime [14] and that LSCO and twinned samples of *YBCO* do not obey Kohler's rule for the magnetoresistance [18]. Nevertheless, the quadratic temperature dependence of the planar resistivity and Kohler's rule clearly are underlying fundamental properties of the $CuO_2$ sheets, with compound-specific deviations that can be understood [14,18,50]. In the present work, we demonstrate for LSCO that the large apparent deviation from the underlying universal behavior of the cotangent of the Hall angle for $x > 0.08$ can be understood to result from a combination of a compound-specific Fermi-surface evolution and effects related to the opening of the PG. Upon correcting for this, the underlying universal value of $C_2$ is recovered. This simple and straightforward result has profound consequences for our understanding of the cuprates. Namely, instead of evoking any exotic interpretations of their physical properties, classic interpretations should be first tried and tested. We thus suspect that the (rather small) deviations from the underlying universal behavior seen in certain compounds, doping and temperature ranges (e.g., Bi2201 [51] in part of the SM region, or Tl2201 at temperatures below 22 K [52]) have a "benign" origin. Just as in the case of LSCO, where these deviations are strongest, such deviations most probably are not the result of a failure of the Fermi-liquid concept. In fact, it has been demonstrated for Tl2201 [25] that a slight difference in Fermi-surface shape may result in a substantially different $\sigma_{xy}$. The temperature- and doping-dependent partial gapping of a non-circular Fermi surface will thus surely cause deviations from the underlying universal transport behavior.

The observation of a FL transport scattering rate in an inherently inhomogeneous system remains a challenge to capture theoretically. It appears necessary to explicitly include the planar oxygen degrees of freedom. We note that calculations indicate that direct oxygen-oxygen propagation may generate arcs with sizeable oxygen spectral weight at the Fermi level [12,30]. Moreover, umklapp interactions seem essential to obtain the $T^2$ scattering rate [12-14]. Regardless of the ultimate microscopic description, the transport results presented here indicate that the transformation from a state with a hole density of $(1+p)/V$ to a state with a density of $p/V$ holes is complete at $T^{**}$ [43,44]. At temperatures below $T^{**}$, these $n_H = p/V$ holes per $CuO_2$ unit give rise to the previously observed "unmasked" FL behavior [14,17,18]. This characteristic temperature furthermore appears to be an upper bound to the charge-density-wave order exhibited by a range of cuprates, including Hg1201 [29,53]. On the very overdoped side of the phase diagram, the observation of pristine FL behavior in both transport and thermodynamic properties [3,4] indicates that the remnants of the PG are gone. Interestingly, the termination of the superconducting phase is concomitant with the disappearance of these remnants. The present work demonstrates that these two FL regions are in fact connected. The largely hidden yet ubiquitous Fermi-liquid scattering rate throughout the cuprate phase diagram and the elucidation of charge transport in the SM region can be expected to enable a more focused discussion of the remaining challenges presented by these fascinating materials.

**Acknowledgements**

N.B. is grateful to the late S. Barišić for extensive discussions. We thank A.V. Chubukov, J. R. Cooper, L. Forró, and D. van der Marel for comments on the manuscript. N.B. acknowledges the support of FWF project P27980-N36. The work at the University of Minnesota was funded by the Department of Energy through the University of Minnesota Center for Quantum Materials, under DE-SC-0016371 and DE-SC-0006858.

## APPENDIX A

**Fermi-liquid (FL) behavior in correlated systems.**

Although the possibility of FL behavior in the PG region of the cuprate phase diagram was initially disputed by many, this interpretation is now increasingly accepted (*e.g.*, refs. [13,54-56]). In fact, there has been a revival of FL concepts in correlated systems beyond the cuprates [57,58]. For example, FL behavior has been found in the iron-pnictides [59,60], and for the Mn-pnictides the evolution from Hund's insulator to a FL was traced in a recent optical spectroscopy study [61]. Even the perfectly compensated semimetal $WTe_2$, which features electron/hole pockets with a Fermi energy of 20-40 meV, exhibits a $T^2$ scattering rate up to high temperatures [62]. A beautiful demonstration of the robustness of FL concepts is the recent work for strontium titanate, which was shown to exhibit a $T^2$ resistivity even at very low carrier densities ($\approx 10^{17}$ cm$^{-3}$) [63]. At those carrier densities, only a single pocket is present at the Fermi level, and the Fermi energy is merely 1 meV.

## APPENDIX B

**Sample preparation and measurement.**

The Hg1201 samples were synthesized, annealed (at various temperatures and oxygen partial pressures in order to adjust the hole concentration) and subsequently contacted according to previously reported procedures [14,16,64]. The superconducting transition temperature ($T_c$) of the samples was determined with a Quantum Design, Inc., Magnetic Properties Measurement System (MPMS). The quoted $T_c$ values correspond to the transition mid-point (10-90% of the diamagnetic signal appears within 1 K). A small "diamagnetic tail" can be found to extend to higher temperatures as a result of filamentary superconductivity. The latter can shorten the sample and cause a resistivity drop already at temperatures above the bulk $T_c$. The doping levels of the samples were assessed using the measured $T_c$ values and the universal thermoelectric power scale [65,66]. The transport measurements were carried out with a Quantum Design, Inc., Physical Properties Measurement System (PPMS). That Hall data were assessed at $H = 9$ T. In order to properly contact *ac* surfaces, the samples had to be cleaved, which introduced an uncertainty of 10-20% in the estimate of the sample contact dimensions [14].

## APPENDIX C

**Resistivity, Hall coefficient and Hall angle in Hg1201.**

Figure 5 shows the temperature dependences of the planar resistivity and Hall effect for eight samples in the doping range $0.05 < p < 0.14$. Representative data for three of these samples are shown in Fig. 2. As discussed in the main text, the resistivity exhibits linear and quadratic temperature dependences above and below the characteristic temperatures $T^*$ and $T^{**}$, respectively. Nevertheless, $\cot(\Theta_H)$ for all samples exhibits doping-independent $T^2$ behavior from just above $T_c$ up to the highest measured temperature of 400 K. We attribute the small differences in the slope $C_2 = \cot(\Theta_H)/T^2$ to the systematic error related to the problematic estimation of the exact sample geometry (Appendix B). Indeed, in Fig. 5c, no systematic tendency can be observed as a function of doping. All Hg1201 samples were found to exhibit negligible extrapolated residual resistance, and thus $\cot(\Theta_H)$ is presented as measured, without any offset.

In order to assess the evolution of the Hall coefficient with doping it is instructive to plot the quantity $eR_Hp/V$, where e is electron charge and $V$ the unit cell volume. This quantity is the ratio of the nominal hole density, $p/V$, to the Hall density, $n_H = 1/(eR_H)$, and should yield a value of 1 if the carrier density measured by the Hall effect corresponds to $p$. As discussed in the main text, for LSCO this is indeed the case in the PG/FL region (for $p = x < 0.08$ and $T < T^{**}$) [24].

Figure 6 shows this ratio as a function of temperature for Hg1201. Below approximately 200 K, all samples exhibit a maximum value that is approximately 20% lower than 1. While there exists an estimated 10-20% uncertainty in the determination of the sample geometry (Appendix B), and hence in the absolute value of $R_H$, the systematically lower value of $eR_Hp/V$ more likely results from a somewhat subtle uncertainty in the estimation of the hole concentration $p$ for those cuprates in which the hole doping level is tuned via changes of interstitial oxygen density [67]. We emphasize that this systematic deviation does not affect any of our conclusions. In particular, we find that $n_H \propto p$, consistent with the result for LSCO ($p = x < 0.08$) and with the observation that the dc-resistivity scales as $1/p$ (ref. [14]).

From Fig. 6 it can also be seen that the peak exhibited by $eR_Hp/V$ is rather broad at lower doping, whereas it is narrower close to optimal doping, consistent with the narrowing of the PG/FL region with increasing hole concentration.

## APPENDIX D

**Analysis of the Hall angle for LSCO ($p > 0.08$).**

As discussed in refs. [28,30,68] and in the main text, with increasing doping the Fermi-surface topology evolves from arcs to a diamond-like shape, and the Hall coefficient ceases to be a good measure of the carrier density. For LSCO, this becomes noticeable already for $p = x > 0.08$. Consequently, $C_2$ begins to deviate from the established universal value (Figs. 3b&7). Nevertheless, the simple quadratic temperature dependence $\cot(\Theta_H) - C_0 \propto T^2$ remains omnipresent. This indicates that the temperature dependence of the evolution of the Fermi surface from arcs toward the full Fermi surface, which is related to gradual closing of the antinodal PG at temperatures above $T^{**}$, is still well captured by the Hall coefficient. However, $R_H$ is related to the curvature of the Fermi surface and ceases to be a reliable measure of the carrier density. Indeed, as shown in Fig. 8, the nodal Fermi-surface curvature ceases to be nearly circular above about $p = 0.08$ and, concomitantly, the Hall effect overestimates the carrier density. The nearly straight Fermi-surface segments near $p \approx 0.20$ result in the apparent divergence of $n_H$.

We consider two methods to correct for this effect. In the first method, we use the recently established temperature and doping dependence of the universal sheet resistance. In contrast to the Hall coefficient, $R_H = \sigma_{xy}/(H\sigma_{xx}\sigma_{yy})$, this observable is not particularly sensitive to the Fermi surface curvature. In the PG/FL part of the phase diagram, the sheet resistance is simply proportional to $T^2/p$ (ref. [14]), and the scattering rate follows $1/\tau \propto T^2$ (refs. [14,17,18]), whereas $R_H$ is approximately constant below $T^{**}$. Since there is very good evidence that the effective mass $m^*$ does not change with doping (e.g., from specific heat [67,69], optical conductivity [19], and dc transport; see main text and ref. [14]), the Drude formula $\rho = m^*/(ne^2\tau)$ implies a carrier density $n = p/V$. This suggests that the Hall data for LSCO ($p > 0.08$) ought to be corrected: $n(T) = (p/V)R_{H,\max}(p)/R_H(p,T)$, where $R_{H,\max}(p)$ is the maximum in the temperature dependence of the Hall coefficient. As demonstrated in Fig. 3b, this adjustment indeed yields the universal value of $C_2$.

The second method to correct $R_H(x,T)$ for $La_{2-x}Sr_xCuO_4$ is very similar to the first. Whereas the first method uses the maximum of $R_H(T)$ at each doping level, the second method is based on ref.

[70], where the carrier density $n(x,T)$ is described in a heuristic fashion as the sum of a doping dependent contribution, $n_0(x)$, that gives the density of carriers in the PG/FL region as measured by the Hall effect, and of a thermally-activated contribution, $n_1(x) \exp[-\Delta(x)/T]$, that is relevant at higher temperatures:

$n_H(x,T) = n_0(x) + n_1(x) \exp[-\Delta(x)/T]$.

Indeed, for $x < 0.08$, $n_0(x)$ equals the nominal carrier concentration $p = x$, whereas for $x > 0.08$ the absolute value of $n_H = 1/(eR_H)$ should be corrected (but not its temperature dependence, as argued above). Instead of using the estimated value $R_{H,max}$, we can use the value of $n_0$ determined from the fit to correct the absolute value of the Hall effect. Due to the rather small difference between in $R_{H,max}$ and $n_0$, the two methods yield very similar results for the corrected value of $C_2$.

## APPENDIX E

**Hall-angle results for multi-layer cuprates.**

$YBa_2Ca_3O_{6+\delta}$ (YBCO) features $CuO_2$ double-layers and CuO chains, and hence is structurally more complex than the single-layer compounds. A quadratic temperature dependence of $\cot(\Theta_H)$ was reported early on up to elevated temperatures ($500^0$ C) [26,71]. Extensive Hall data were obtained in subsequent work, and an attempt was made to separate contributions from the $CuO_2$ planes from those of the CuO chains [72]. These latter data, which span a rather wide doping range and are reproduced in Figs. 9&10, demonstrate that, for $p < 0.10$, $C_2$ is consistent with the universal value established for the single-layer compounds. For $p > 0.10$, YBCO behaves similar to LSCO. Figure 10 also shows data for double-layer $Bi_2SrCa_{1-x}Y_xCu_2O_{8+\delta}$ (Bi2212; ref. [73]) and triple-layer $HgBa_2Ca_2Cu_3O_{8+\delta}$ (Hg1223; ref. [74]).

The Fermi surfaces of these compounds are inherently more complicated due to the multi-layer structure. As noted, in the case of YBCO it is necessary to carefully separate CuO chain from $CuO_2$ plane contributions. In the Bi-based cuprates, the underlying quadratic temperature dependence of the resistivity in the PG/FL region is masked (ref. [14]), yet the quadratic scattering rate is clearly revealed in $\cot(\Theta_H)$. In triple-layer Hg1223, the inequivalent inner and outer layers might have different hole concentrations and different levels of intrinsic inhomogeneity. In light of this additional complexity, the data for the multi-layer cuprates are remarkably consistent with those for the single-layer compounds discussed in the main text.

## APPENDIX F

**General observations related to the optical conductivity and photoemission spectroscopy.**

Recent analysis of the zero-frequency limit of the optical conductivity established for several transition-metal oxides the temperature dependence of the effective plasma frequency and revealed a hidden FL scattering rate, $1/\tau \propto T^2$ (ref. [57]). This result was obtained for $V_2O_3$ and $NdNiO_3$, two correlated materials that exhibit a temperature-driven metal-insulator transition, and for $CaRuO_3$, which is a Hund's metal. The observation that the anomalous transport properties of these complex materials arise from a temperature-dependent plasma frequency ($\omega_p^2 \propto n/m^*$) is consistent with our findings, yet the temperature dependence was ascribed to the effective mass rather than the carrier density [57]. We can exclude the former possibility in the case of the cuprates, since $\cot(\Theta_H) - C_0 = C_2T^2 \propto m^*/\tau$ is universal, and thus very unlikely the result of a compensation between separate doping/temperature dependences of the scattering rate and effective mass. Our result instead naturally demonstrates the existence of a universal transport scattering rate, $1/\tau \propto T^2$, throughout the entire phase diagram, and it suggests that the effective mass is independent of doping and temperature.

This conclusion finds further support from the comparison in Fig. 11 between the temperature dependences of the Hall coefficient, $R_H = 1/(en_H)$, and the ratio of optical effective mass and carrier density, $m^*/n$, for Hg1201. The latter ratio was determined at an energy of 10 meV through an extended Drude analysis and originally attributed to a temperature-dependent effective mass [17,75]. However, we emphasize that the optical conductivity measures $m^*/n$, and not $m^*$ itself [76]. The good agreement between the temperature dependences of $R_H \propto 1/n_H$ and $m^*/n$ therefore supports our conclusion. Nevertheless, one needs to keep in mind the limits of the extended Drude analysis, which assumes frequency dependences of the scattering rate and of the effective mass, and a constant charge-carrier density. In the case of the cuprates, it seems necessary for the frequency and temperature dependences to be treated on equal footing. In this respect, our result for the cotangent of the Hall angle provides a key constraint (see Appendix G).

We note that quantum oscillation experiments for underdoped YBCO and Hg1201 indicate a small, pocket-like Fermi-surface with an effective mass of $m^* = 2\text{-}3\ m_e$ ($m_e$ is the free-electron mass) that is approximately half as large as that for highly doped Tl2201 [77]. Whereas the small Fermi surface measured in underdoped Hg1201 and YBCO is the result of a Fermi-surface reconstruction [29,77] that can be expected to modify the effective mass, the experiments on Tl2201 determine the large unreconstructed Fermi surface [4] and hence are directly relevant to the present discussion.

Photoemission experiments are typically carried out on the bismuth-based cuprates for which neither $\rho \propto T^2$ nor Kohler scaling has been reported in the PG/FL region. This discrepancy could be the result of the particularly high degree of inhomogeneity exhibited by these compounds [36,37], which might cause a substantial temperature dependence of the carrier density even at temperatures below $T^{**}$. And indeed, despite the apparent non-FL behavior of the resistivity, $\cot(\Theta_H)$ for double-layer $Bi_2SrCa_{1-x}Y_xCu_2O_{8+\delta}$ exhibits a quadratic temperature dependence with a constant $C_2$ that agrees rather well with the universal value (Appendix E). The single-particle lifetime measured in photoemission might differ from the transport lifetime. Nevertheless, for LSCO ($p = x = 0.23$), photoemission yields a quadratic frequency dependence of the self-energy in the nodal region, consistent with our findings, but an approximately linear behavior in the antinodal region [34]. At the same doping level the resistivity can be decomposed into quadratic and linear terms [14] and $R_H(T)$ still exhibits significant temperature dependence below about 200 K [24,31]. Consequently, in those parts of the phase diagram where $R_H(T)$ exhibits a substantial temperature dependence and where, in accord with our finding of a universal scattering rate, the carrier density can not be assumed to be independent of temperature, it seems necessary for photoemission results in the antinodal region to be re-analyzed.

## APPENDIX G

**Suggestion of a novel approach to analyze the optical conductivity.**

In the main text, we briefly discuss optical conductivity data in order to demonstrate that these (as well as photoemission) results are not inconsistent with our proposed interpretation of the dc transport data. Our main points are: (i) the extended Drude analysis reveals the Fermi-liquid scattering rate ($1/\tau$) for temperatures below $T^{**}$, in agreement with the dc transport data; (ii) the temperature dependence of $m^*/n$ resembles the evolution of $n$ determined from the Hall effect (Fig. 11).

We emphasize that the commonly applied extended Drude analysis should be treated with care in the case of systems for which a simple single-band approach is insufficient. This seems to be particularly relevant to the cuprates at temperatures above $T^{**}$, i.e., outside the PG/FL region

More generally, in light of our dc transport results, we believe that all optical conductivity data for the cuprates should be re-analyzed. While this analysis is clearly beyond the scope of the present paper, our transport results give several key constraints. First, instead of the prior approaches, we suggest that the low-frequency data that correspond to the coherence peak are fit throughout the entire phase diagram (i.e., not only in the PG/FL and FL, but also in the SM) in a Fermi-liquid manner, with $1/\tau \propto (\hbar\omega)^2 + (a\,\pi k_B T)^2$. We suggest three different approaches to fix the coefficient $a$: (i) from fits to $1/\tau \propto (\hbar\omega)^2 + (a\,\pi k_B T)^2$ in the PG/FL or FL regions; (ii) from an extended Drude analysis in the PG/FL or FL regions (e.g., $a = 1.5$ [17]); (iii) $a$ can be fixed to the theoretically predicted value $a = 2$ [78,79]. Once the value of $a$ has been determined, it should be held fixed in the SM region, and the coherent spectral weight should be matched to the carrier density determined from the Hall effect. This procedure implies that there is essentially no free parameter for the coherent spectral weight in the SM region.

The second constraint pertains to the delocalization of one carrier per planar copper atom upon increasing the temperature above $T^{**}$. The redistribution of spectral weight with temperature at a fixed nominal doping level $p$ should bear important similarities to the evolution with doping at a fixed temperature. Indeed, it is remarkable that the spectral weight of the coherent contribution at temperatures below $T^{**}$ corresponds to $p$, whereas the spectral weight integrated to about 2.5 eV (on the order of the charge-transfer gap) approaches $1 + p$ (where we attribute 1 to the localized charge). With this in mind, one can now take a fresh look at the doping dependence of the optical conductivity in this energy range. We consider the extensive early data for LSCO [80] which are reproduced in Fig. 12. Without any detailed analysis, it is indeed possible to follow the evolution of this charge from a hump at intermediate energy at low doping ($x = p$ for LSCO) to the quasiparticle peak at high doping. We expect a corresponding evolution with temperature at fixed doping.

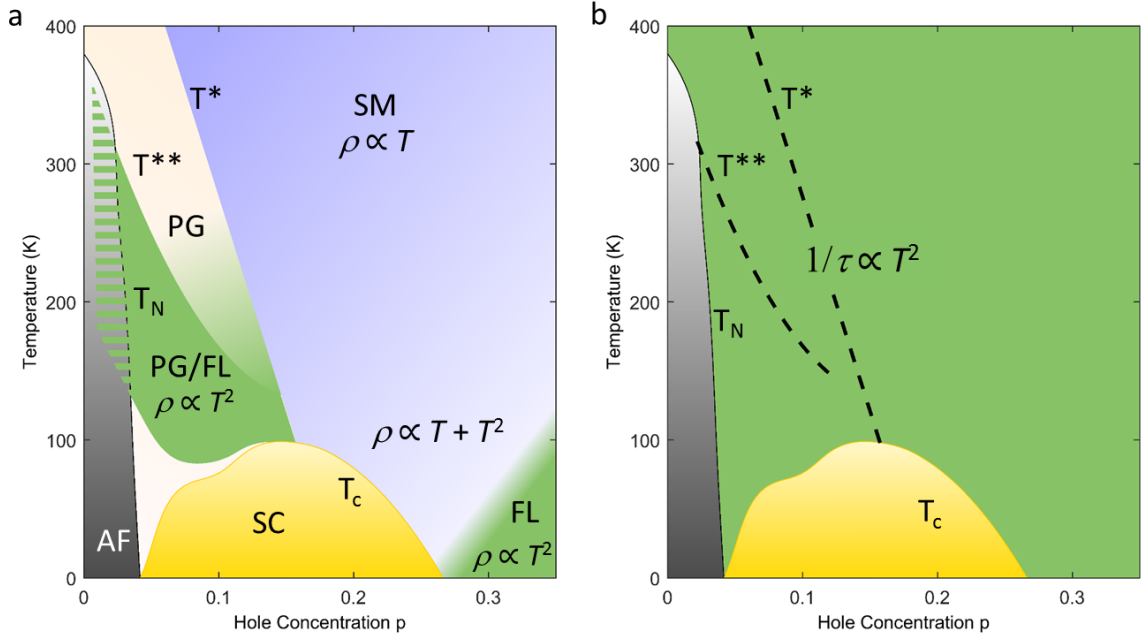

**Figure 1. Schematic phase diagram**. **(a)** Phase diagram with emphasis on resistivity results [14]. Conventional metallic Fermi-liquid (FL) $\rho \propto T^2$ behavior is observed on both sides of the superconducting (SC) phase. At zero and low hole-dopant concentrations, antiferromagnetic (AF) order is observed below the Néel temperature $T_N$. At temperatures above the pseudogap (PG) temperature $T^*$, the planar resistivity exhibits an approximately linear temperature dependence, $\rho \propto T$. This part of the phase diagram is often referred to as a strange metal (SM). Below the characteristic temperature $T^{**}$, FL behavior (including $\rho \propto T^2$) is observed even half-way into the AF phase at $p = 0.01$ [14,24]. This behavior terminates at low temperatures due to SC fluctuations ($p > 0.05$) or disorder effects ($p < 0.05$). **(b)** Consideration of the Hall angle yields the transport scattering rate and reveals the underlying FL behavior across the entire phase diagram. After accounting for compound-specific issues, the underlying behavior $1/\tau \propto T^2$ is seen to be universal, i.e., doping and compound independent (see also Fig. 3 and Appendix D & E). The two seemingly disjointed FL states in **a** are therefore de facto connected, and the $T$-linear resistivity appears to be the result of a temperature-dependent carrier density. Approximate temperature/doping dependences of $T^*$ and $T^{**}$ are indicated by dashed lines.

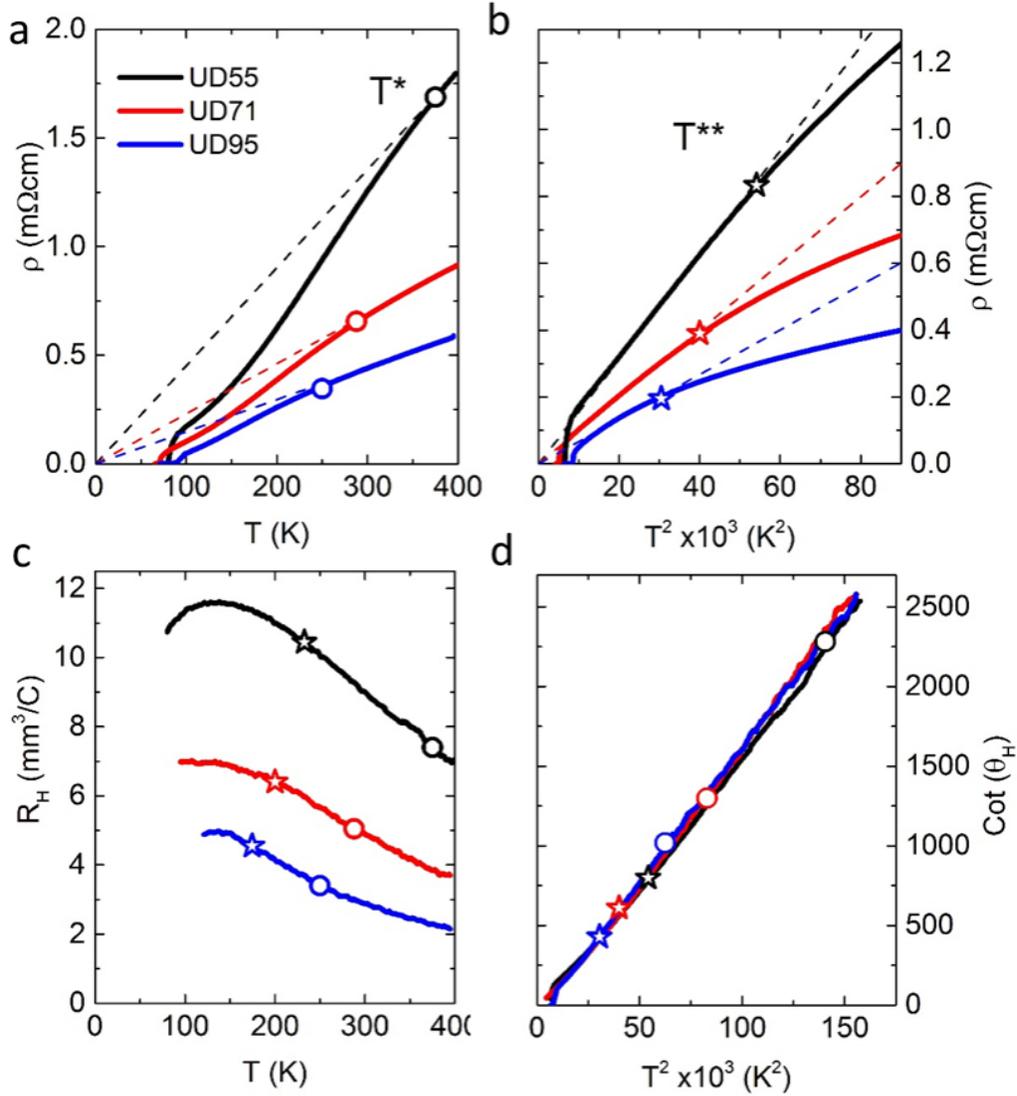

**Figure 2. Temperature dependence of resistivity, Hall coefficient and cotangent of Hall angle in Hg1201.** Planar resistivity as a function of (**a**) $T$ and (**b**) $T^2$ for three representative samples with $T_c$= 55 K, 71 K and 95 K (see also Appendix C - Fig. 4 shows data at additional doping levels). Deviation from $\rho \propto T$ indicates the PG temperature $T^*$ (circles), whereas $T^{**}$ (stars) marks the onset of the $\rho \propto T^2$ region. (**c**) Hall coefficient $R_H$ for the same samples, with characteristic temperatures indicated. $T^{**}$ corresponds to an inflection point. In the temperature range $T_c + 15$ K $< T < T^{**}$, $R_H$ is approximately constant for moderately doped samples. (**d**) In contrast to $\rho$ and $R_H$, $\cot(\Theta_H)$ is doping independent. Unlike for the resistivity, the quadratic temperature dependence of $\cot(\Theta_H)$ is unchanged across $T^{**}$ and $T^*$.

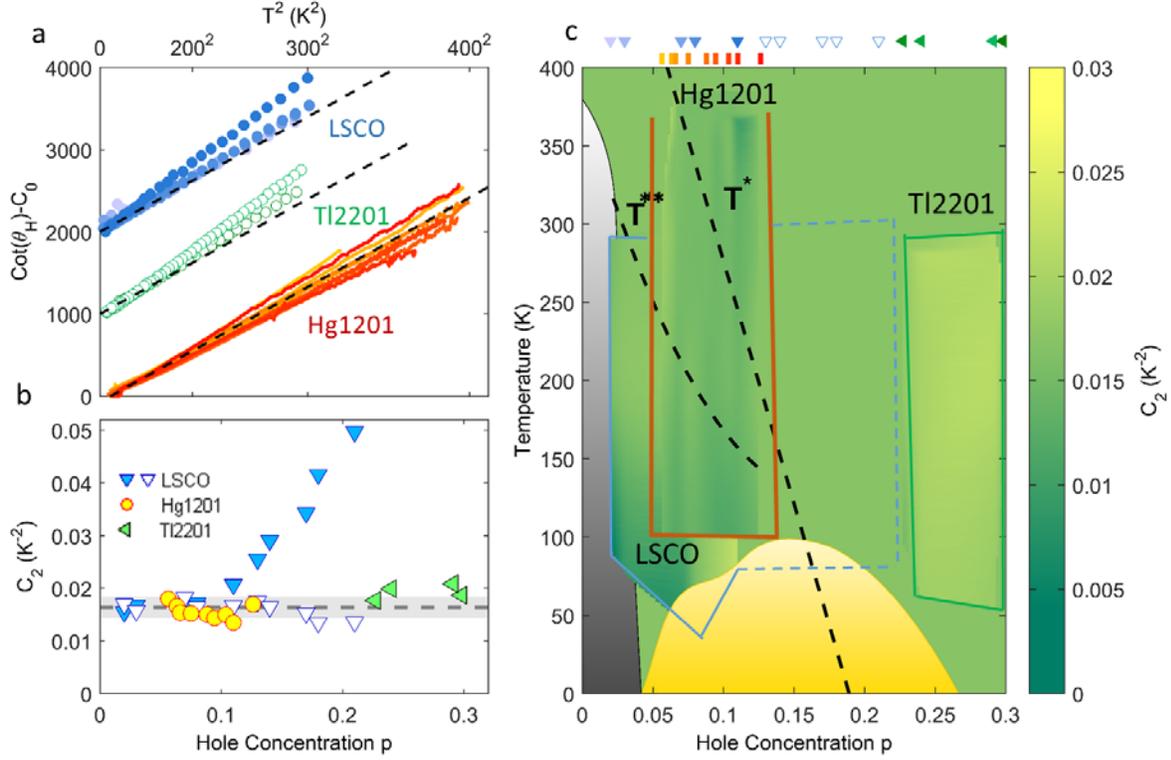

**Figure 3. Universal transport scattering rate. (a)** Temperature dependences of $\cot(\Theta_H)$ for LSCO, Tl2201 and Hg1201. The Hg1201 data exhibit a negligible residual contribution ($C_0$) and are plotted as measured. This contribution appears to be related to disorder and is subtracted for LSCO and Tl2201. For clarity, the LSCO and Tl2201 data are then shifted vertically by 1000 and 2000 units, respectively. Data for LSCO are shown up to $p = x = 0.10$. The universal $\cot(\Theta_H) \propto T^2$ dependence is clearly seen. The small scatter in the Hg1201 data can be ascribed to the uncertainty in assessing sample dimensions (see Appendix C); a similar uncertainty likely exists for the LSCO and Tl2201 data as well. **(b)** Doping dependence of $(\cot(\Theta_H) - C_0)/T^2$ demonstrates the universal value of $C_2$ throughout the phase diagram (see also Appendix D&E). Dashed line and shaded grey area indicate the mean value and standard deviation of $C_2 = 0.0175(20)$ K$^{-2}$, respectively, obtained from fits to all the data except for LSCO at doping levels above $p = 0.08$. For LSCO, the data above $p = x = 0.08$ deviate from the universal value as the system undergoes a Fermi-surface topology change [28]. The open blue triangles for LSCO are the corrected values, as detailed in the text and in Appendix D. **(c)** The universal $C_2T^2$ behavior and underlying transport scattering rate is found throughout the phase diagram. AF and SC phases as well as $T^*$ and $T^{**}$ are indicated as in Fig. 1. The measured temperature/doping ranges are indicated for all three compounds by highlighted areas (including the extended dashed frame for the corrected LSCO result) and by symbols that correspond to those in **a**. Background color corresponds to the mean value of $C_2 = 0.0175$ in the right contour scale.

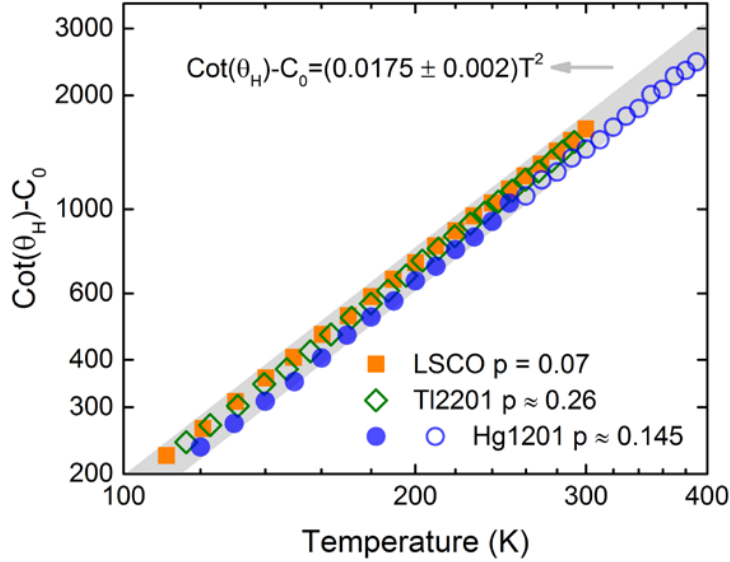

**Figure 4. Cotangent of the Hall angle for three single-layer cuprates.** The temperature dependences of $\cot(\Theta_H) - C_0$ for LSCO, Tl2201 and Hg1201 are compared on a log-log scale in order to highlight the remarkable universal (doping and compound independent) behavior. The LSCO data (solid squares, from Ref. [24]) correspond to $T < T^*$, the Tl2201 data (open squares, from Ref. [25]) lie well outside the PG region, whereas the Hg1201 data span both regions: $T > T^*$ (open circles) and $T < T^*$ (filled circles). The gray band indicates the range $C_2 = 0.0175$ +/- 0.002 (same as in Fig. 3b).

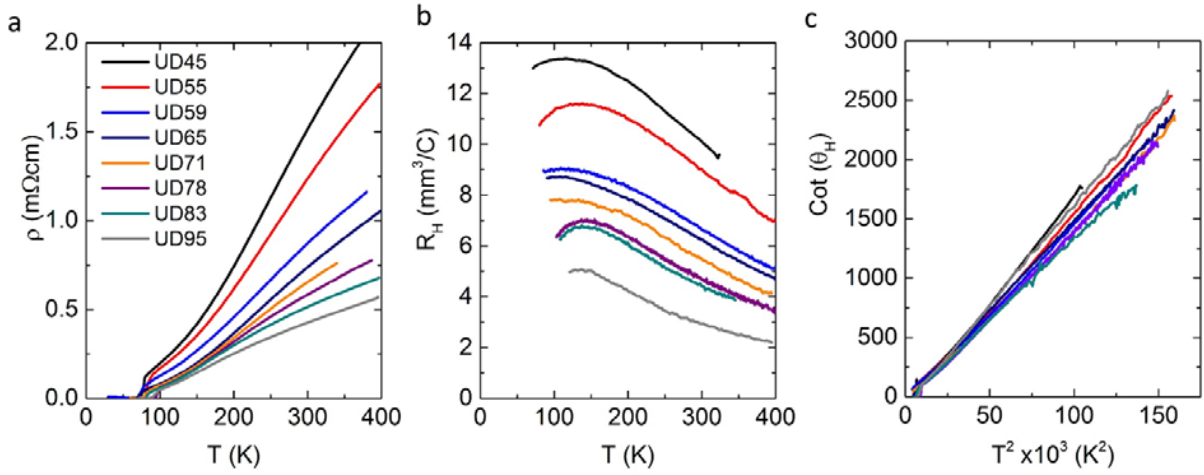

**Figure 5. Temperature dependencies of resistivity, Hall coefficient and Hall angle for eight Hg1201 samples.** **(a)** Temperature dependence of the planar resistivity of samples that range from very underdoped ($T_c = 45$ K) to nearly optimally doped ($T_c = 95$ K). **(b)** Hall coefficient for the same samples. **(c)** $\cot(\Theta_H)$ determined from the data in **a** and **b** (at $H = 9$ T).

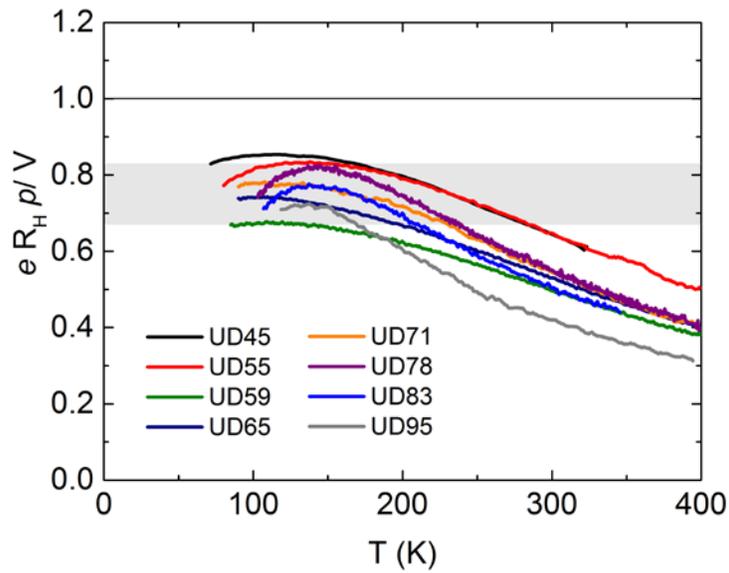

**Figure 6. Nominal carrier concentration versus Hall effect.** Temperature dependence of the ratio of nominal hole concentration, $p/V$, and measured carrier density, $n_H = 1/(eR_H)$. The grey horizontal band indicates the range 0.80 +/- 0.08.

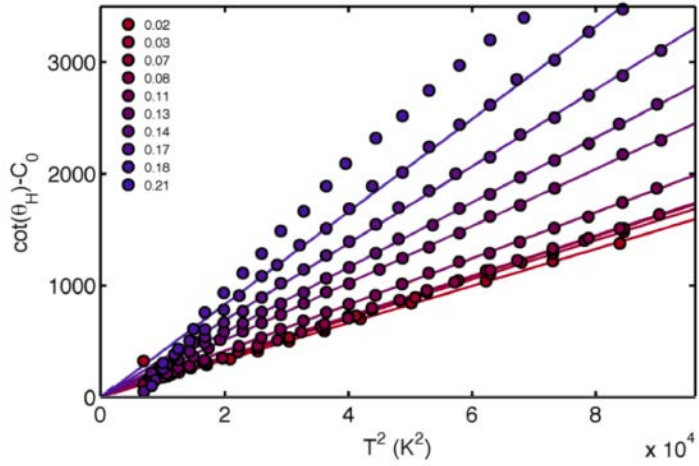

**Figure 7. Hall angle in LSCO for $p > 0.08$.** Temperature dependence of $\cot(\Theta_H) - C_0$ for LSCO shows quadratic behavior for a very wide range of Sr doping values $x$ (data from ref. [24]). Low-temperature data are not shown for clarity, due to large upturns for the most underdoped samples. Solid lines are fits to $C_2 T^2$. The so-obtained $C_2$ values are shown in Fig. 3b. The data for $x = 0.21$ do not exhibit a quadratic temperature dependence and were not fit.

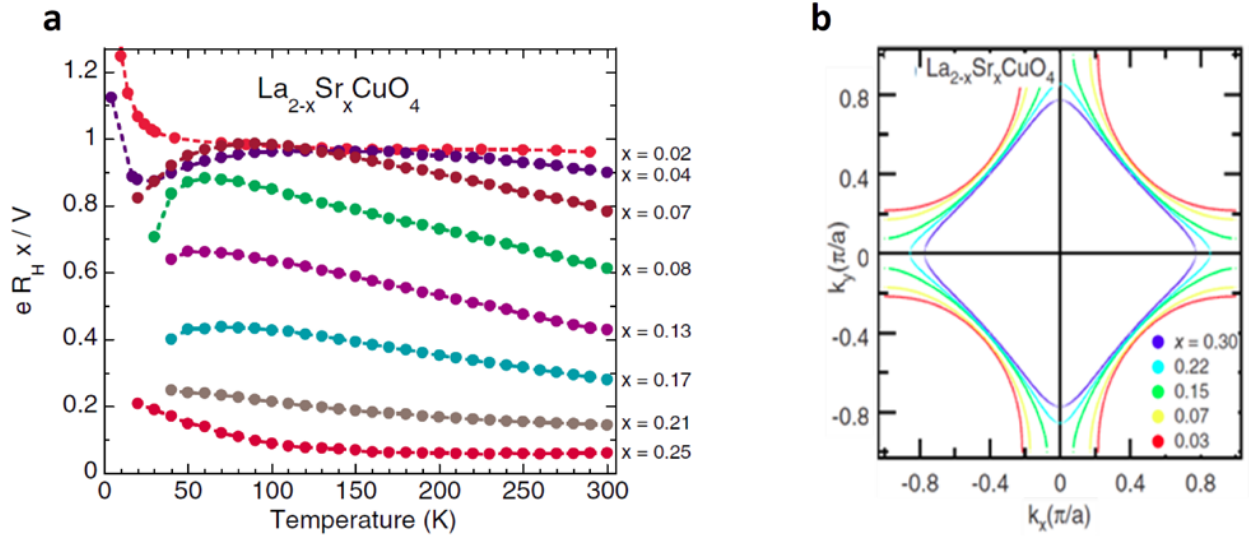

**Figure 8. Comparison of the doping evolution of the Hall coefficient $R_H(T)$ and the underlying Fermi surface of LSCO.** (**a**) Temperature dependences of $eR_Hx/V$, the ratio of the measured Hall coefficient to that expected from nominal hole density, $x/V$, for a wide range of doping levels. For $x < 0.08$, this ratio remains close to one in the measured temperature range (the Hall density agrees with the nominal carrier density), and the planar resistivity exhibits quadratic temperature dependence [24]. (**b**) Tight-binding description of the Fermi surface of LSCO [28,68], consistent with experimental data. Below $x = 0.08$, the underlying Fermi surface is approximately circular. Upon further doping, the Fermi-surface shape increasingly deviates from a circular shape and becomes nearly 'flat' in the nodal direction, which causes an overestimation of the carrier density measured via the Hall effect. Consequently, for $x > 0.08$, the ratio $eR_Hx/V$ begins to decrease from 1. In addition, $R_H$ exhibits substantial temperature dependence already below 300 K and the maximum value moves to lower temperature. **a** and **b** were adopted from Ref. [24] and Refs. [28,68], respectively.

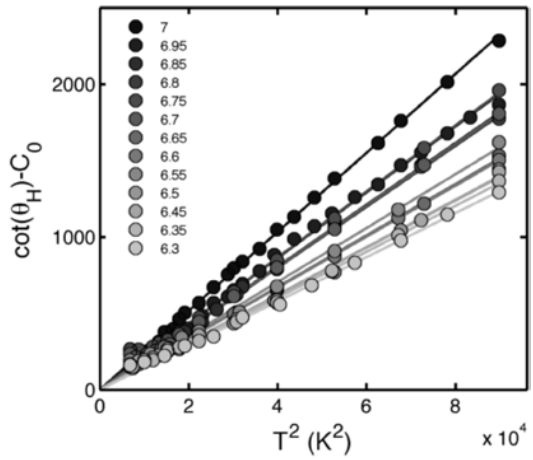

**Figure 9. Doping evolution of the Hall angle in YBCO.** $\cot(\Theta_H) - C_0$ for $YBa_2Cu_3O_y$ for samples with different oxygen content $y$, from Ref. [72]. Data at low temperatures (below 80 K) are not shown for clarity, due to large upturns for the most underdoped samples. Solid lines are fits to the form $C_2T^2$.

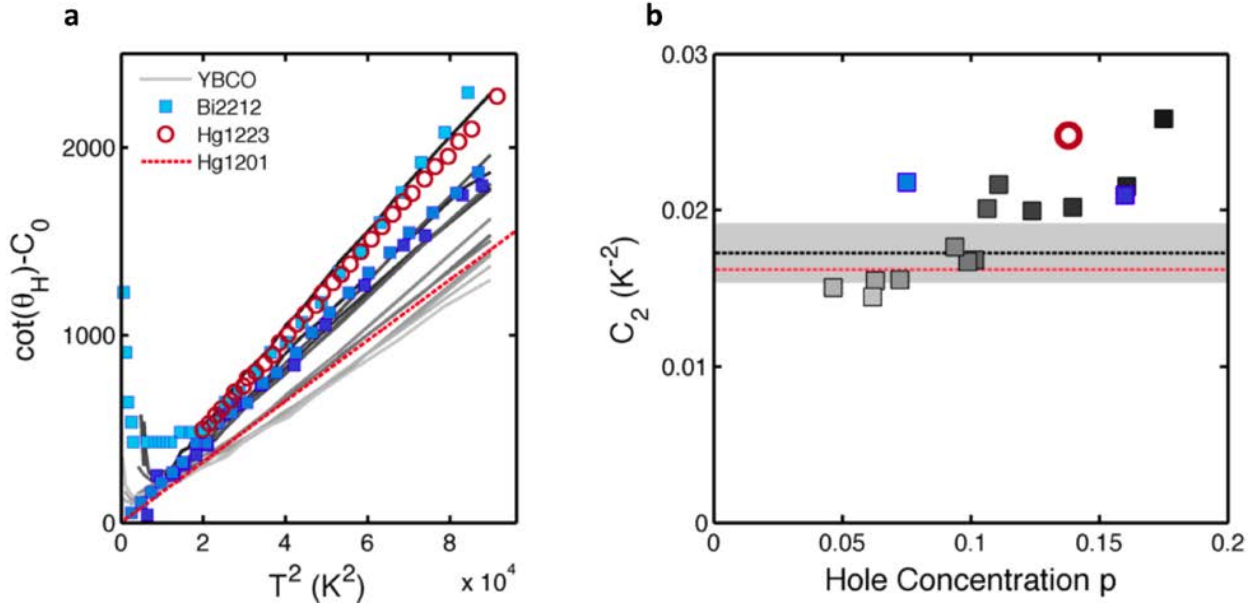

**Figure 10. Hall angle for multi-layer cuprates. (a)** $\cot(\Theta_H) - C_0$ as a function of $T^2$ for YBCO (grey lines as in Fig. 7 – data from Ref. [72]), Bi2212 ($T_c$ = 96 K, optimally doped, dark blue squares; $T_c$ = 40 K, underdoped, medium blue squares; non-superconducting, very underdoped, light blue squares – data from Ref. [73]) and Hg1223 ($T_c$ = 134 K, nearly optimally doped, red circles – data from Ref. [74]). Also shown is the average result for Hg1201 obtained in the present work (dotted red line). Except for YBCO below $p$ = 0.10, the slope of the underlying quadratic temperature dependence is considerably larger than the established underlying universal value, consistent with the situation for LSCO near optimal doping (Appendix D). **(b)** $C_2$ as a function of doping for YBCO, superconducting Bi2212, and Hg1223. The doping levels were estimated based on the Obertelli-Cooper-Tallon formula [65] and the quoted values of $T_c$. The dotted line and grey shaded band are the same as in **a**.

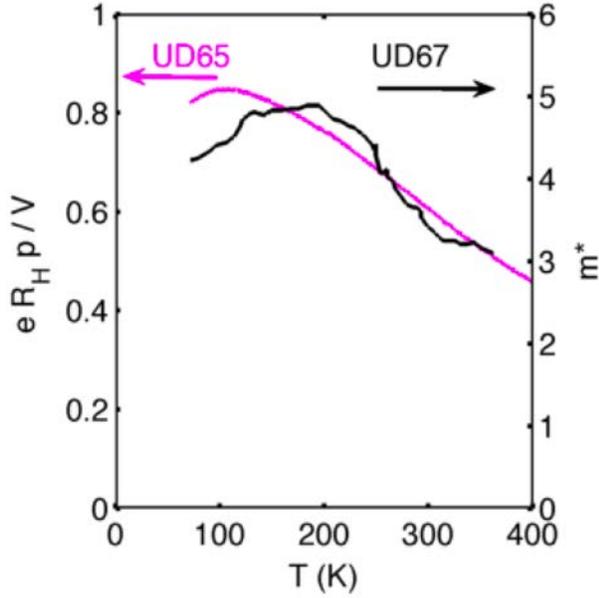

**Figure 11. Comparison with optical conductivity result.** Comparison for Hg1201 of the normalized Hall coefficient with $m^*$ (in units of the electron mass) at 10 meV, determined in Ref. [17] from the optical conductivity (samples with $T_c$ = 65 and 67 K, respectively). In Ref. [17], the temperature dependence was attributed to $m^*$, although, in effect, the measured quantity is $m^*/n$. The good correspondence between the temperature dependences of $R_H \propto 1/n_H$ and $m^*/n$ suggests the opposite: the carrier density exhibits a substantial temperature dependence above $T^{**} \approx 200$ K, whereas the effective mass is independent of temperature.

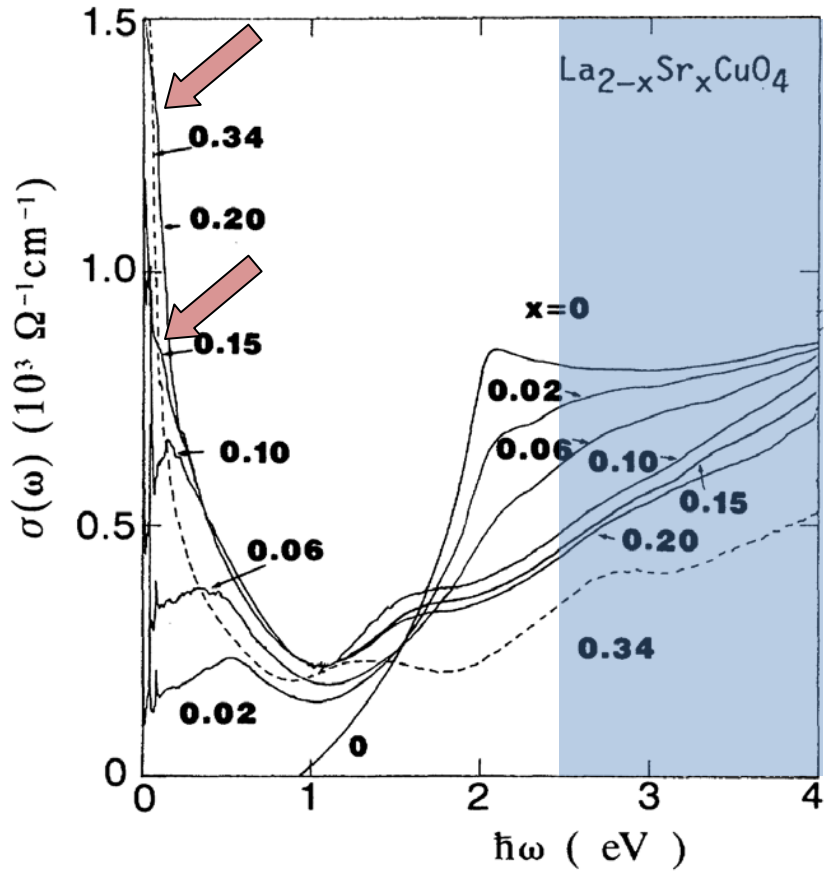

**Figure 12. Planar optical conductivity for LSCO, measured at room temperature (adapted from ref. 80).** The frequency range of interest (up to about 2.5 eV; white background) approximately corresponds to $1 + p$ holes per $CuO_2$ unit [17]. We associate the "1" with the doping-dependent "hump" that moves toward low energy with increasing nominal hole concentration $p$. At low doping, this feature is easily seen as a local maximum. At $x = p = 0.15$ and 0.20, the hump is indicated with arrows. At $x = 0.15$, the hump is seen to have nearly merged with the coherent quasiparticle peak, whereas at $x = 0.20$ only a weak "shoulder" can be discerned (arrows). Finally, at $x = 0.34$, the low-frequency response is dominated by a simple quasiparticle peak.